\documentstyle[eclepsf]{l-aa}

\begin{document}

   \thesaurus{06         
              (02.18.8;  
               08.13.1;  
               08.14.1;  
               03.13.1)} 
   \title{Deformation of
       relativistic magnetized stars}


   \author{K. Konno \and T. Obata \and Y. Kojima
          }

   \offprints{K. Konno}

   \institute{Department of Physics, 
              Hiroshima University,
              Higashi-Hiroshima 739-8526, Japan
             }

   \date{Received }

   \maketitle

   \begin{abstract}

   We formulate deformation of relativistic stars 
   due to the magnetic stress, considering the magnetic fields
   to be perturbations from spherical stars.
   The ellipticity for the dipole magnetic field is calculated
   for some stellar models. 
   We have found that the ellipticity becomes large 
   with increase of a relativistic factor for the  
   models with the same energy ratio
   of the magnetic energy to the gravitational energy.

   \end{abstract}

\section{Introduction}

There is a growing interest in new classes of objects: 
soft-gamma repeaters (SGRs) and anomalous X-ray pulsars (AXPs).
Until now, four identified SGRs are reported observationally.
Their associations with supernova remnants strongly suggest
that the SGRs are young neutron stars 
(see e.g. Kulkarni \& Frail \cite{kf};
Murakami et al. \cite{murakami}).
Furthermore, recent measurements 
(see e.g. Kouveliotou et al. \cite{kouv1}, \cite{kouv2}) of 
the period and period derivative yield evidence for these
pulsars to be ultramagnetized neutron stars with field
strength ($\sim 10^{15} \mbox{G}$)
in excess of $B_{\rm cr} \sim 10^{13} \mbox{G}$,
i.e., {\em magnetars} (Duncan \& Thompson \cite{dt92};
Thompson \& Duncan \cite{td93,td95,td96}).
Some class of X-ray pulsars  
also suggests the 
magnetic fields with $10^{14}$--$10^{15} \mbox{G}$
(see e.g. Mereghetti \& Stella \cite{ms}).
Such magnetic fields are much stronger than that of 
known pulsars ($10^{8}$--$10^{13}\mbox{G}$;
see e.g. Taylor et al. \cite{catalog}) until then.
Though the relation between the SGRs and the AXPs is not yet clear,
there exist neutron stars with very strong magnetic fields.
In these ultramagnetized stars, the magnetic influence
becomes important as well as the relativistic effects. 
If we assume that a long-lived electric 
current flows in highly conductive neutron-star matter, the magnetic
pressure corresponding to the Lorentz force comes into
play. Hence, it induces deformation of stars.
In this paper, we study such deformation from spherical stars
within a general relativistic framework.

The quadrupole deformation of magnetized Newtonian stars 
was discussed by Chandrasekhar \& Fermi (\cite{cf}) and 
Ferraro (\cite{ferraro}), in which incompressible fluid body 
with a dipole magnetic field is assumed.
This deformation has been discussed also in relation to
the gravitational radiation (Gal'tsov et al. \cite{gtt};
Gal'tsov \& Tsetkov \cite{gt}).
The general relativistic approach by Bocquet et al.~(\cite{bbgn})
and Bonazzola \& Gourgoulhon (\cite{bg})
has appeared recently. 
However, their approach is fully numerical. 
In this paper, we develop almost an analytical treatment
by assuming weak magnetic fields compared with gravity.
This assumption is valid even in the magnetars.
Our formulation
is regarded as a general relativistic version
of Chandrasekhar \& Fermi (\cite{cf}) and Ferraro (\cite{ferraro}).
In our method, we can easily include realistic equations of state
(EOS) and construct relativistic magnetized stars.
Furthermore, this method gives simple calculations of ellipticity of 
deformed stars, and so on.

Since the observed ultramagnetized neutron stars have long
periods ($T \sim \mbox{several sec}$),
we may neglect the rotation of the magnetized stars, that is,
we discuss static cases.
We take non-rotating, spherical relativistic stars as 
backgrounds, and consider the magnetic fields as the perturbation.
In particular, we consider only axisymmetric,
poloidal magnetic fields produced by long-lived 
(toroidal) electric currents, because toroidal
magnetic fields would break the symmetric property
(see also Bocquet et al.~\cite{bbgn} and reference therein).
Furthermore, we assume a perfectly conducting
interior. Since we now consider non-rotating
configurations, this implies that the electric field
inside the stars must be zero.
Hence, there is no electric charge inside the stars. 
From this, we can write the 4-current as 
$J_{\mu} = \left( 0, 0, 0, J_{\phi} \right)$
(Bocquet et al.~\cite{bbgn}).
Furthermore, the surface charge should be absent,
since the total charge should vanish in astrophysical 
situations. Otherwise, the electromagnetic field itself 
would have the angular momentum due to the non-vanishing 
electric field produced by the charge
(Feynman et al.~\cite{feynman};
Ma \cite{ma}; de Castro \cite{castro}).
This is not the purely static case.

The current distribution is introduced as the first-order
quantity with respect to the perturbation. 
The corresponding magnetic field is solved 
by the Maxwell equation.
We shall investigate deformation of stars
due to the resultant magnetic stress, which arises as 
the second-order correction to the background field.
This perturbation method is very similar to that of 
slowly rotating stars developed by Hartle (\cite{hartle}),
in which the rotation is regarded as a small parameter.
Our formalism can be applied to any configurations 
of the magnetic fields.
However, we restrict ourselves to dipole magnetic fields
because the dipole fields are important in 
many astrophysical situations.

The plan of this paper is as follows.
In Sect.~2, the magnetic fields are investigated in the background
space-time. The effect arising from the magnetic
stress on equilibrium of stars is considered in Sect.~3.
The solution corresponding to the quadrupole deformation of the stars 
will be given. To evaluate the deformation quantitatively, 
the ellipticity is discussed in Sect.~4. Finally, Sect.~5
is devoted to the discussion.
Throughout this paper, we use the units in which $c=G=1$, 
and the Gaussian system of units for electromagnetic fields.

\section{Magnetic fields in spherically
symmetric space-time}

We now consider an axisymmetric, poloidal magnetic field 
created by a 4-current $J_{\mu} = (0,0,0,J_{\phi})$ 
in a non-rotating, spherical star. We suppose that the magnetic
field is weak, i.e., $B_{\mu} \sim O( \varepsilon )$. 
The line element of the spherically symmetric space-time is 
given by
\begin{equation}
 ds^2 = - e^{\nu (r)} dt^2 + e^{\lambda (r)} dr^2 
            + r^2 d \theta^2 + r^2 \sin^2 \theta d \phi^2 ,
\end{equation}
where $\nu$ and $\lambda$ are functions of the radial
coordinate $r$ only.
Since we do not consider electric fields,
we can assume that the electromagnetic 4-potential $A_{\mu}$
has only the $\phi$-component, i.e., $A_{\mu} = (0,0,0,A_{\phi})$
(see also Bocquet et al.~\cite{bbgn}). 
In this case, the Maxwell equation is reduced to
\begin{eqnarray}
\label{la}
 \lefteqn{e^{- \lambda} \frac{\partial^2 A_{\phi}}{\partial r^2}
 + \frac{1}{2} \left( \nu' - \lambda' \right) 
   e^{-\lambda} \frac{\partial A_{\phi}}{\partial r}}
        \nonumber \\
 & & + \: \frac{1}{r^2} \frac{\partial^2 A_{\phi}}{\partial \theta^2}
  - \frac{1}{r^2} \cot \theta \frac{\partial A_{\phi}}{\partial \theta}
  \; = \; - 4 \pi J_{\phi} ,
\end{eqnarray}
where the prime denotes the differentiation with respect to $r$.

We expand the potential $A_{\phi}$ and the current $J_{\phi}$
as follows (Regge \& Wheeler \cite{rw}):
\begin{eqnarray}
\label{a-expansion}
 A_{\phi} & = & \sum_{l=1}^{\infty} a_{l} (r) 
                \sin \theta \frac{d P_{l}(\cos \theta)}{d \theta} , \\
\label{j-expansion}
 J_{\phi} & = & \sum_{l=1}^{\infty} j_{l} (r) 
                \sin \theta \frac{d P_{l}(\cos \theta)}{d \theta} ,
\end{eqnarray}
where $P_{l}$ is the Legendre's polynomial of degree $l$.
Substituting these forms into Eq.~(\ref{la}), we have
\begin{equation}
\label{difeq_a1}
 e^{-\lambda} \frac{d^2 a_{l}}{dr^2}
 + \frac{1}{2} \left( \nu' - \lambda' \right) e^{-\lambda}
   \frac{d a_{l}}{dr} - \frac{l(l+1)}{r^2} a_{l} 
 = - 4 \pi j_{l} .
\end{equation}
For a given current $j_{l}$, we can obtain the potential 
$a_{l}$ and, therefore, the magnetic field.
From now on, we only consider a dipole magnetic field, i.e.,
$l=1$. The potential outside the star is easily solved
(Ginzburg \& Ozerno\u\i \, \cite{go};
Petterson \cite{petter}; Wasserman \& Shapiro \cite{ws})
in the form
\begin{equation}
\label{ext_a}
 a_{1} = - \frac{3 \mu}{8M^3} r^2 \left[ \ln \left( 1 - \frac{2M}{r} \right)
       + \frac{2M}{r} + \frac{2M^2}{r^2} \right] ,
\end{equation} 
where $\mu$ is a constant corresponding to the magnetic 
dipole moment with respect to an observer at infinity, and
$M$ is the total mass of the background star.
In order to describe the magnetic field inside 
the star, we require the current distribution $j_{1}$.
The current $j_{1}$ is not arbitrary but
subject to an integrability condition 
(Ferraro \cite{ferraro}; 
Chandrasekhar \& Prendergast \cite{cp};
Bonazzola et al. \cite{bgsm}).
As will be shown in Eq.~(\ref{current-d}), this current is given,
up to the first order in $\varepsilon$, by 
\begin{equation}
 j_{1} (r) = c_{0} r^2 \left( \rho_{0}(r) + p_{0}(r) \right) ,
\end{equation}
where $c_{0}$ is an arbitrary constant, and $\rho_{0}$ and
$p_{0}$ denote the density and pressure, respectively, 
of the background star. By requiring that $a_{1}$
behaves as a regular function at the center of the star,
we now obtain the potential $a_{1}$ in the vicinity of the center:
\begin{eqnarray}
\label{a_ext}
 a_{1} & \simeq & \alpha_{0} r^2 
    + O\left( r^4 \right) ,
\end{eqnarray}
where $\alpha_0$ is a constant, which is fixed by the 
boundary condition at the surface.
In this way, we can construct the magnetic field in the whole 
space-time.

   \begin{figure}[htbp]
      \epsfile{file=9149.f1,width=8.8cm}
      \caption{The tetrad component of the magnetic fields,
    $B_{\hat{r}}$ on the symmetry axis ($\theta=0$),
    plotted against $r/R$.
    The solid line denotes a relativistic case ($M/R=0.2$), 
    while the dashed line corresponds to a Newtonian case 
    ($M/R=0.01$). The magnetic fields are normalized by the
    typical field strength $\mu / R^3$.} 
    \label{mgField-Br}
   \end{figure}
   \begin{figure}[htbp]
      \epsfile{file=9149.f2,width=8.8cm}
      \caption{The tetrad component of the magnetic fields,
    $B_{\hat{\theta}}$ on the equatorial plane 
    ($\theta=\pi /2$), plotted against $r/R$.
    The solid line denotes a relativistic case ($M/R=0.2$), 
    while the dashed line corresponds to a Newtonian case 
    ($M/R=0.01$). The magnetic fields are normalized by the
    typical field strength $\mu / R^3$.}
     \label{mgField-Bth}
   \end{figure}

Fig.~\ref{mgField-Br} displays the tetrad component of 
the magnetic field,
\begin{equation}
  B_{\hat{r}} = -\frac{1}{r^2 \sin \theta} 
         {\partial}_{\theta}A_{\phi} 
       = \frac{2 \cos \theta}{r^2} a_1 
\end{equation} 
on the symmetry axis (i.e., $\theta =0$),
and Fig.~\ref{mgField-Bth} displays the tetrad component 
\begin{equation}
  B_{\hat{\theta}} = \frac{e^{- \frac{\lambda}{2}}}{r \sin \theta} 
         {\partial}_r A_{\phi}
       = -\frac{e^{- \frac{\lambda}{2}} \sin \theta}{r} a'_1 
\end{equation} 
on the equatorial plane (i.e., $\theta = \pi /2$)
with respect to the radial coordinate $r$. 
We have normalized them by the typical 
magnetic field strength $\mu / R^3$, where $R$ is the radius
of the star.
The solid lines denote a relativistic case, whereas
the dashed lines correspond to a Newtonian case.
In these calculations, we have used the polytropic
EOS: $p_0= \kappa \rho_0^{\gamma} \; (\gamma=2)$.
From these figures, we see that the intensity 
of the magnetic field increases as $r$ becomes closer 
to the center.
Furthermore, these two figures show that despite the same magnetic moment
with respect to an observer at infinity, the central 
magnetic field of the relativistic star is stronger 
than that of the Newtonian star by about 50\% of the 
Newtonian case. 
Therefore, it follows that the relativistic effect 
strengthens the internal magnetic fields.

\section{Equilibrium configurations of magnetized stars}

Next, we consider deformation of magnetized stars due to 
the magnetic stress, which is regarded as the
second-order effect. We formulate the deformation of the 
star and space-time, following Hartle (\cite{hartle}).

\subsection{Equations of equilibrium}

The metric can be expanded in multipoles around the
spherically symmetric space-time. In particular, when 
we deal only with a dipole field, i.e., $l=1$ in 
Eqs.~(\ref{a-expansion}) and (\ref{j-expansion}),
the metric can be written in the form 
(see also Hartle \cite{hartle};
Chandrasekhar \& Miller \cite{cm})
\begin{eqnarray}
 ds^2 & = & - e^{\nu (r)} \left[ 1 + 2 \left( h_{0}(r) + h_{2}(r) 
          P_{2}( \cos \theta ) \right) \right] dt^2  \nonumber \\
  & & + \: e^{\lambda (r)} \left[ 1 + \frac{2 e^{\lambda (r)}}{r} 
          \left( m_{0}(r) + m_{2}(r) P_{2}( \cos \theta )
          \right) \right] dr^2 
          \nonumber \\
  & & + \: r^2 \left[ 1 + 2 k_{2}(r) P_{2}( \cos \theta ) 
        \right] \left( d \theta^2 
        + r^2 \sin^2 \theta d \phi^2 \right) \!,
\end{eqnarray}
where $h_0$, $h_2$, $m_0$, $m_2$ and $k_2$ are the corrections
of the second order in $\varepsilon$.

The total energy-momentum tensor
is the sum of the perfect-fluid part 
$T^{\quad \mu}_{^{\rm (m)}\ \nu}$ and the electromagnetic part
$T^{\quad \mu}_{^{\rm (em)}\ \nu}$:
\begin{equation}
 T^{\mu}_{\ \nu} 
    = T^{\quad \mu}_{^{\rm (m)}\ \nu} 
      + T^{\quad \; \mu}_{^{\rm (em)}\ \nu} ,
\end{equation}
where 
\begin{eqnarray}
 T^{\quad \mu}_{^{\rm (m)}\ \nu} 
 & = & \left( \rho + p \right) u^{\mu} u_{\nu} + p \delta^{\mu}_{\ \nu} , \\
\label{em-T}
 T^{\quad \; \mu}_{^{\rm (em)}\ \nu} 
 & = & \frac{1}{4\pi} \left( F^{\mu \lambda}
     F_{\nu \lambda} - \frac{1}{4} F_{\sigma \lambda}
     F^{\sigma \lambda} \delta^{\mu}_{\ \nu} \right) .
\end{eqnarray}
In Eq.~(\ref{em-T}), $F_{\mu \nu}$ is the Faraday tensor.
The pressure $p$ and the energy density $\rho$ can also 
be expanded in multipoles as 
\begin{eqnarray}
\label{pressure}
 p(r,\theta) 
 & = & p_{0} + \left( \delta p_{(l=0)}
       + \delta p_{(l=2)} P_{2} \right), \\
 \rho(r,\theta) 
 & = & \rho_{0} + \frac{\rho_{0}'}{p_{0}'}
       \left( \delta p_{(l=0)} + \delta p_{(l=2)} 
       P_{2} \right) ,
\end{eqnarray}
where $\delta p_{(l=0)}$ and $\delta p_{(l=2)}$
depend on $r$ only, and we have assumed a barotropic case.

From the Einstein equation, we can obtain 
\begin{eqnarray}
 m_{0}' 
 = 4 \pi r^2 \frac{\rho_{0}'}{p_{0}'} \delta p_{(l=0)}
       + \frac{1}{3} \left[ e^{-\lambda} \left( a_{1}' \right)^2
       + \frac{2}{r^2} a_{1}^{2} \right] , 
\end{eqnarray}
\begin{eqnarray}
 h_{0}' 
 & = & 4 \pi r e^{\lambda} \delta p_{(l=0)}
       + \frac{1}{r} \nu' e^{\lambda} m_{0} 
       + \frac{1}{r^2} e^{\lambda} m_{0} \nonumber \\
 & & + \: \frac{e^{\lambda}}{3r} \left[ e^{-\lambda} 
       \left( a_{1}' \right)^2
       - \frac{2}{r^2} a_{1}^{2} \right] , 
\end{eqnarray}
\begin{eqnarray}
 h_{2}' + k_{2}' 
 & = & h_{2} \left( \frac{1}{r} - \frac{\nu'}{2} \right)
       + \frac{e^{\lambda}}{r} m_{2} \left( \frac{1}{r} 
       + \frac{\nu'}{2} \right) \nonumber \\
 & & + \: \frac{4}{3r^2} a_{1} a_{1}' , 
\end{eqnarray}
\begin{eqnarray}
\label{aq-h2m2}
 h_{2} + \frac{e^{\lambda}}{r} m_{2}
 = \frac{2}{3} e^{- \lambda} \left( a_{1}' \right)^2 , 
\end{eqnarray}
\begin{eqnarray}
 \lefteqn{h_{2}' + k_{2}' + \frac{1}{2} r \nu' k_{2}'}
   \nonumber \\
 & & = 4 \pi r e^{\lambda} \delta p_{(l=2)}
       + \frac{1}{r^2} e^{\lambda} m_{2}
       + \frac{1}{r} \nu' e^{\lambda} m_{2} 
       \nonumber \\
 & & \quad + \: \frac{3}{r} e^{\lambda} h_{2}
       + \frac{2}{r} e^{\lambda} k_{2} 
       - \frac{1}{3r} e^{\lambda} \left[ 
         e^{-\lambda} \left( a_{1}' \right)^2 
       + \frac{4}{r^2} a_{1}^{2} \right] .
\end{eqnarray}
Furthermore, from the conservation law of the total energy-momentum
tensor, we obtain
\begin{eqnarray}
\label{p20d}
 \delta p_{(l=0)}' 
 & = & - \frac{1}{2} \nu' \left( \frac{\rho_{0}'}{p_{0}'}
       + 1 \right) \delta p_{(l=0)}  \nonumber \\
 & & - \: \left( \rho_{0} + p_{0}
         \right) h_{0}' + \frac{2}{3r^2} a_{1}' j_{1} , \\
\label{p22}
 \delta p_{(l=2)} 
 & = & - \left( \rho_{0} + p_{0} \right) h_{2}
       - \frac{2}{3r^2} a_{1} j_{1} , \\
\label{p22d}
  \delta p_{(l=2)}'
 & = & - \frac{1}{2} \nu' \left( \frac{\rho_{0}'}
         {p_{0}'} + 1 \right) \delta p_{(l=2)} 
       \nonumber \\
 & & - \: \left( \rho_{0} + p_{0} \right) h_{2}'
       - \frac{2}{3r^{2}} a_{1}' j_{1} .
\end{eqnarray}
The integrability condition for Eqs.~(\ref{p20d}) and (\ref{p22d})
leads to
\begin{equation}
\label{current-d}
 \frac{j_{1}}{r^2 \left( \rho_{0} + p_{0} \right)}
   = \mbox{const.} \left( \equiv c_{0} \right) .
\end{equation}
This is consistent with Eq.~(5.29) of Bonazzola et al.~(\cite{bgsm}) 
up to the first order in $\varepsilon$.
Using this current distribution, Eq.~(\ref{p20d}) can 
also be integrated as
\begin{equation}
\label{small-p20}
  \delta p_{(l=0)}  
  = - \left( \rho_0 + p_0 \right) h_0 + \frac{2}{3 r^{2}}
    a_1 j_1 + c_1 \left( \rho_{0} + p_{0} \right),
\end{equation}
where $c_{1}$ is a constant of integration.

Consequently, we have two set of differential equations,
\begin{eqnarray}
\label{m0h0-1}
 m_0' & = & -4 \pi r^{2}\frac{\rho_0'}{p_0'}
    \left( \rho_0 +\ p_0 \right) \left( h_0 -c_1 \right) 
    \nonumber \\
 & & + \: \frac{1}{3} e^{-\lambda} \left( a_1' \right)^2
    + \frac{2}{3 r^{2}} a_1^{2} 
    + \frac{8 \pi}{3} \frac{\rho_0'}{p_0'}
    a_1 j_1 , \\
\label{m0h0-2}
 h_0' & = & \left( \frac{1}{r^2} + \frac{\nu'}{r} \right) 
      e^{\lambda} m_0 - 4 \pi r e^{\lambda} 
      \left( \rho_0 +p_0 \right) \left( h_0 - c_1 \right)
      \nonumber \\
 & & + \: \frac{1}{3 r} \left( a_1' \right)^{2} 
     - \frac{2}{3 r^3} e^{\lambda} {a_1}^{2}
     + \frac{8 \pi}{3 r} e^{\lambda} a_1 j_1,
\end{eqnarray}
and
\begin{eqnarray}
\label{v2h2-1}
 v_2'& = & - \nu' h_2 + \frac{2}{3} e^{- \lambda}
      \left( \frac{1}{r} + \frac{\nu'}{2} \right)
      \left( a_1' \right)^2
      + \frac{4}{3 r^2} a_1 a_1' , \\
\label{v2h2-2}
 h_2' & = & - \frac{4e^{\lambda}}{r^2 \nu'} v_2
     + \left[ 8 \pi \frac{e^{\lambda}}{\nu'} 
     \left( \rho_0 + p_0 \right)
     +  \frac{2}{r^2 \nu'} \left( 1 - e^{\lambda} \right)
     - \nu' \right] h_2  \nonumber \\
 & & + \: \frac{8}{3 r^4 \nu'} e^{\lambda} a_1^2  
     + \frac{8}{3 r^3 \nu'} \left( 1 + \frac{r \nu'}{2} \right) a_1 a_1'
     \nonumber \\
 & & + \left( \frac{1}{3} \nu' e^{- \lambda} 
     + \frac{2}{3 r^2 \nu'} \right) \left( a_1' \right)^2
     + \frac{16 \pi}{3 r^2 \nu'} e^{\lambda} a_1 j_1 ,
\end{eqnarray}
where $v_{2} \equiv h_{2} + k_{2}$.
These equations govern the relativistic magnetized star.
We have to solve four differential equations 
(\ref{m0h0-1})--(\ref{v2h2-2}) and one algebraic 
equation (\ref{aq-h2m2}) for the unknown functions
$(m_{0}, m_{2}, h_{0}, h_{2}, v_{2})$.
Furthermore, we can derive $\delta p_{(l=0)}$
and $\delta p_{(l=2)}$ by substituting the solution 
of $h_{0}$ and $h_{2}$ into Eqs.~(\ref{small-p20}) and (\ref{p22}).

In order to solve Eqs.~(\ref{m0h0-1}) and (\ref{m0h0-2})
inside the star, it is also convenient to introduce a quantity
\begin{equation}
 \delta P_{0} \equiv \frac{\delta p_{(l=0)}}{\rho_{0}+p_{0}} .
\end{equation}
From Eq.~(\ref{small-p20}), we have
\begin{equation}
 \delta P_{0} + h_0 - \frac{2}{3} 
   \frac{j_1}{r^2 \left(\rho_0 + p_0 \right)} a_1 = c_1 .
\end{equation}
Moreover, Eqs.~(\ref{m0h0-1}) and (\ref{m0h0-2}) are rewritten as
\begin{eqnarray}
\label{m0p0-1}
 m_0' & = & 4 \pi r^2 \frac{\rho_0'}{p_0'} 
   \left( \rho_0 + p_0 \right) \delta P_0
   + \frac{1}{3} e^{-\lambda} \left( a_1' \right)^2
   + \frac{2}{3 r^{2}} a_1^{2} , \\
\label{m0p0-2}
 \delta P_0' & = & - \left( 8 \pi p_0 + \frac{1}{r^2} \right)
   e^{2 \lambda} m_0 - 4 \pi r e^{\lambda} 
      \left( \rho_0 +p_0 \right) \delta P_0
      \nonumber \\
 & & - \: \frac{1}{3 r} \left( a_1' \right)^{2} 
     + \frac{2}{3 r^3} e^{\lambda} {a_1}^{2}
     + \frac{2}{3} \frac{j_1}{r^2 \left( \rho_0 + p_0 \right)}
     a_1' .
\end{eqnarray}
In the next subsection, we solve these differential 
equations for the metric functions.

\subsection{The exterior solution and boundary condition}

First, we consider the solution 
outside of the star, in which $\rho_0 = p_0 =0$ and $j_1 = 0$.

The solution of $m_{0}$ and $h_{0}$ is given by
\begin{eqnarray}
 m_0 & = & \frac{3 \mu^2}{8 M^4 r} \left( r^2 - M^2 \right)
           \nonumber \\
 & & + \: \frac{3 \mu^2}{8 M^5} \left( r^2 - M r - M^2 \right)
   \ln \left( 1- \frac{2M}{r} \right) \nonumber \\
 & & + \: \frac{3 \mu^2}{32 M^6} r^2 \left( r - 2M \right)
   \left[ \ln \left( 1- \frac{2M}{r} \right) \right]^2 + c_2 , \\
 h_0 & = & - \: \frac{3 \mu^2}{8 M^3} 
   \frac{4r-M}{r \left( r-2M \right)} 
   \nonumber \\
 & & + \: \frac{3 \mu^2}{8 M^5}
   \frac{(r-M)(r-3M)}{r-2M} \ln \left( 1 - \frac{2M}{r} \right)
   \nonumber \\
 & & + \: \frac{3 \mu^2}{32 M^6} r^2 \left[ \ln
   \left( 1 - \frac{2M}{r} \right) \right]^2
   - \frac{c_2}{r-2M} + c_3 ,
\end{eqnarray}
where $c_2$ and $c_3$ are constants of integration. 
At large $r$, $m_0$ and $h_0$ behave as
\begin{eqnarray}
 m_0 & \simeq & c_2 - \frac{\mu^2}{3 r^3} 
   + O \left( \frac{1}{r^4} \right), \\
 h_0 & \simeq & c_3 - \frac{3 \mu^2}{8 M^4} - \frac{c_2}{r}
   + O \left( \frac{1}{r^2} \right) .
\end{eqnarray}
Since $h_0$ must vanish at infinity, we obtain
\begin{equation}
 c_3 = \frac{3 \mu^2}{8 M^4} .
\end{equation}
The constant $c_{2}$ corresponds to the mass shift, which
is fixed by matching with the interior solution
at the surface.

On the other hand, the differential equations for $v_2$ and $h_2$
is rather complicated, but analytically solved.
The solution of $v_2$ is
\begin{equation}
\label{v2-ext}
 v_{2} (z) = \frac{2K}{\sqrt{z^2-1}} Q^1_2 (z)
   - \frac{3 \mu^2}{4M^4 \sqrt{z^2-1}} P^1_2 (z) + v_{2p} (z), \;
\end{equation}
where $z$ is defined as $z \equiv r/M-1$, $K$ is a constant of
integration, $P^1_2$ and $Q^1_2$ are the associated Legendre functions,
and $v_{2p}$ is 
\begin{eqnarray}
 v_{2p} & = & \frac{9 \mu^2}{4 M^4} z
   + \frac{3 \mu^2}{8 M^4} \frac{7 z^2 - 4}{z^2 - 1}
     \nonumber \\
 & & + \: \frac{3 \mu^2}{16 M^4} \frac{z \left( 11 z^2 - 7 \right)}
   {z^2 - 1} \ln \frac{z-1}{z+1} \nonumber \\
 & & + \: \frac{3 \mu^2}{16 M^4}
   \left( 2 z^2 + 1 \right) \left( \ln \frac{z-1}{z+1} \right)^2 .
\end{eqnarray}
Furthermore, $h_2$ is given by
\begin{equation}
\label{h2-ext}
 h_2(z) = K Q^2_2 (z) - \frac{3 \mu^2}{8 M^4} P^2_2 (z) +h_{2p} (z) ,
\end{equation}
where $P^2_2$ and $Q^2_2$ are the associated Legendre functions,
and $h_{2p}$ is written as
\begin{eqnarray}
 h_{2p} & = & - \: \frac{3 \mu^2}{16 M^4} \left[
   6 z^2 + 3 z - 6 - \frac{4z^2+2z}{z^2-1} \right]
   \nonumber \\
 & & - \: \frac{3 \mu^2}{32 M^4} \left[ 3 z^2
   - 8 z - 3 - \frac{8}{z^2 - 1} \right] \ln \frac{z-1}{z+1} 
   \nonumber \\
 & & + \: \frac{3 \mu^2}{16 M^4} \left( z^2 - 1 \right) \left(
   \ln \frac{z-1}{z+1} \right)^2 .
\end{eqnarray}
In Eqs.~(\ref{v2-ext}) and (\ref{h2-ext}), we have 
used the boundary condition at infinity.
The remaining constant $K$ will be fixed by the 
boundary condition at the surface.
Thus we have obtained the analytical solution outside the star.

We turn our attention to the interior solution.
For a given EOS, we can obtain the solution numerically.
For the actual numerical work, 
we investigate the behavior of the metric functions
in the vicinity of the center.

First, we consider the metric functions $m_0$ and $\delta P_0$.
The solution in which both $m_0$ and $\delta P_0$ vanish
at the center (see also Chandrasekhar \& Miller \cite{cm}) 
is given by 
\begin{eqnarray}
 m_0 & \simeq & \frac{2}{3} \alpha_0^{\ 2} r^3 
   + \cdots , \\
 \delta P_0 & \simeq & - \frac{2}{3} \left(
   \alpha_0^{\ 2} - c_0 \alpha_0 \right) r^2
   + \cdots ,
\end{eqnarray}
where $\alpha_0$ is defined in Eq.~(\ref{a_ext}).

Next, we consider $v_{2}$ and $h_{2}$. 
The regular solution at the center is
\begin{eqnarray}
 v_{2} & \simeq & \beta_{1} r^4 + \cdots ,  \nonumber \\
 h_{2} & \simeq & \beta_{2} r^2  + \cdots ,
\end{eqnarray}
where constants $\beta_{1}$ and $\beta_{2}$ are not 
independent by the regularity condition
at the center.

Finally, we can obtain the metric functions by
imposing the junction conditions
(O'Brien \& Synge \cite{os}) at the surface:
\begin{eqnarray}
 \left. g_{\mu \nu} \right|_{+R} 
   & = & \left. g_{\mu \nu} \right|_{-R}
 \quad \left( \mu, \nu = t, r, \theta , \phi \right) , \\
 \left. g_{ij, r} \right|_{+R} 
   & = & \left. g_{ij, r} \right|_{-R}
 \quad \left( i , j = t, \theta , \phi \right) ,
\end{eqnarray}
where $g_{\mu \nu}$ denotes the metric components.
From these conditions, the integration constants 
$c_{2}$, $K$, $\beta_{1}$ and $\beta_{2}$ are fixed.

\section{Ellipticity of magnetized stars}

We consider the magnetic field on the stellar shape
of the equilibrium. The additional Lorentz force
$\vec{J} \times \vec{B}$ mainly acts on it in the perpendicular
direction to the symmetry axis
($\theta = \pi /2$), that is, flattens the star.
The flattening effect is also recognized by 
considering the $(r,r)$ component
of the magnetic stress tensor,
\begin{equation}
 T_{^{\rm (em)} \ r}^{\quad \; r}
  = \frac{1}{8 \pi} \left( B_{\theta} B^{\theta}
     - B_{r} B^{r} \right) .
\end{equation}
Along the symmetry axis, $B_{\theta}$ must vanish owing to
the axisymmetry. Hence, the stress $T_{^{\rm (em)} \ r}^{\quad \; r}$
is negative on this axis. On the other hand, on the
equatorial plane, since $B_{r}$ is zero at any $r$, 
the stress $T_{^{\rm (em)} \ r}^{\quad \; r}$ has the opposite sign.
This indicates that the spherical star is shrunk in the 
parallel direction to the symmetry axis ($\theta = 0$)
and expanded in the perpendicular direction ($\theta = \pi /2$) 
by the magnetic effect.
Thus we can see the flattening effect.

Next, in order to evaluate the deformation quantitatively,
let us introduce the ellipticity, 
which is defined as
\begin{equation}
  \mbox{ellipticity} \equiv  
  \frac{(\mbox{equatorial radius})-( \mbox{polar radius})}
    {(\mbox{mean radius})} ,
\end{equation}
where these radii denote the circumferential radii
under general relativistic situations. 
From this definition, ellipticity is given by
(Chandrasekhar \& Miller \cite{cm})
\begin{eqnarray}
 \label{elp}
 \mbox{ellipticity}
  & = & \frac{2c_0}{r\nu'} a_1 +  \frac{3h_2}{r \nu'}
        -\frac{3}{2}k_2 .       
\end{eqnarray}
The first term of the right-hand side in Eq.~(\ref{elp}) 
corresponds to, in a sense, the effect of the `Lorentz force',
the second term represents the effect of
the perturbation of the `gravitational 
potential' induced by the magnetic effect,
and the third term is a `purely
relativistic term' which arises from a definition of 
the radius, that is, the circumferential radius.
These contributions to the ellipticity are shown 
in Fig.~\ref{elp-t} as a function of the relativistic factor $M/R$.
We have normalized them by $\mu^2 R^2 /I^2$
($I$ is the moment of inertia, which is well defined 
also in the relativistic case), and we have used
a polytropic EOS ($\gamma=2$). 
From this figure, we can see that the term due to the 
`gravitational potential' does not change significantly,
while the term concerning the `Lorentz force' increases
with the relativistic factor $M/R$
as known from Figs.~\ref{mgField-Br} and \ref{mgField-Bth}.

   \begin{figure}[htbp]
      \epsfile{file=9149.f3,width=8.5cm}
      \caption{The three contributions to ellipticity:
    (a) the effect of the `Lorentz force',
    (b) the effect of the perturbation of the `gravitational 
    potential', and (c) the `purely relativistic effect' (see text).
    These are plotted against $M/R$.
    We have used the polytropic EOS ($\gamma = 2$).}
      \label{elp-t}
   \end{figure}
   \begin{figure}[htbp]
      \epsfile{file=9149.f4,width=8.5cm}
      \caption{The variation of the ellipticity with respect to $M/R$.
    The solid line denotes the case of the polytropic EOS with $\gamma=2$, 
    and the dashed line corresponds to the case of $\gamma=5/3$.}
    \label{elp-MR}
   \end{figure}

Fig.~\ref{elp-MR} displays the (total) ellipticity for different
polytropic models ($\gamma = 5/3$ and $\gamma = 2$).
From this figure, we find that the ellipticity becomes large
as the relativistic factor $M/R$ increases, in each case
of $\gamma = 5/3$ and $\gamma = 2$.
The common feature of the monotonic increase
shows the effect of the 
`Lorentz force term' to be effective.
An important thing is that the relativistic calculation leads to
much larger ellipticity for a fixed value of 
$\mu^2 R^2 / I^2$.
Finally, we give a comment concerning
the previous general-relativistic studies.
The quantity plotted in Fig.~\ref{elp-MR} is exactly the 
{\em magnetic distortion factor} $\beta$ introduced
by Bonazzola \& Gourgoulhon (\cite{bg}).
It is noticed that although the EOS which we have used is different from 
that of Bonazzola \& Gourgoulhon (\cite{bg}), 
both calculations give the same results (see Fig.~2 of their paper),
that is, $\beta$ takes the values very close to $1$.

\section{Discussion}

Recent observations suggest that some class of neutron stars
has strong magnetic fields. These may promote a new branch
with magnetized stars. As a simple approach to the models,
we have formulated the structure of the magnetized stars
within a general relativistic framework, 
considering the perturbation from a spherical star. 
In particular, a dipole magnetic field has been dealt with. 
We have showed the current distribution which 
yields equilibrium configurations
up to the second order in $\varepsilon$.
Furthermore, the ellipticity of the star has been estimated
as a simple example.
We have found that the ellipticity becomes large
as the relativistic factor $M/R$ increases, for the same energy ratio
of the magnetic energy to the gravitational energy.
Our analytical approach have made
the calculations much simpler than that of the
previous work.
This method can be extended to more general cases
of realistic EOS and general current
distribution, in which the current exists in some domain
of the star. Therefore, this can be applied to 
wider range of astrophysical situations.

Another extension of this work is to incorporate rotation of stars.
The stationary configurations, in which 
the rotation axis is aligned with the magnetic axis,
make the calculations complex because of appearance
of non-vanishing electric fields. However, this can be managed. 
Since the rotational effect deforms the star as well as
the magnetic effect, we are also interested in seeing
which of them to be effective.
These will be the subject of further investigation.

\begin{acknowledgements}
   We would like to thank M.~Hosonuma for providing us with 
   a numerical code for rotating stars and useful discussions.
\end{acknowledgements}

\end{document}